\documentclass[final,5p,times,twocolumn]{elsarticle}

\usepackage{graphicx} 
\usepackage{xcolor} 
\usepackage{epstopdf}  
\usepackage{amsmath,bm} 
\usepackage{enumerate}
\usepackage{float}
\usepackage{tablefootnote}

\newtheorem{assum}{Assumption}

\usepackage{multirow}
\usepackage[hidelinks]{hyperref}

\newcommand{\myfrac}[2]{%
	\ifinner#1/#2%
	\else\frac{#1}{#2}%
	\fi%
}

\newcommand{\p}{\text{p}}
\usepackage{amssymb}

\journal{Journal of Process Control}

\begin{document}

\begin{frontmatter}
 
   \title{A predictive modular approach to constraint satisfaction under uncertainty - with application to  glycosylation in continuous monoclonal antibody biosimilar production}  

\author[label1,label2]{Yu Wang
}
\address[label1]{Division of Decision and Control Systems, KTH Royal Institute of Technology, Stockholm 10044, Sweden}
\address[label2]{Competence Centre for Advanced Bioproduction by Continuous Processing, AdBIOPRO, Sweden\fnref{label4}}

\cortext[cor1]{Corresponding author}

\ead{wang3@kth.se}
\author[label1]{Xiao Chen}
\ead{xiao2@kth.se}

\author[label2,label4]{Hubert Schwarz}
\ead{hschwarz@kth.se}

\author[label2,label4]{V\'{e}ronique Chotteau}
\ead{chotteau@kth.se}
\address[label4]{ Division of Industrial Biotechnology,  KTH Royal Institute of Technology, Stockholm 10044, Sweden}

\author[label1,label2]{Elling W. Jacobsen\corref{cor1}}
\ead{jacobsen@kth.se}

\begin{abstract}

The paper proposes a modular-based approach to constraint handling in process optimization and control. This is partly motivated 
by the recent interest in learning-based methods, e.g.,  within bioproduction, for which constraint handling under uncertainty is 
a challenge. The proposed constraint handler, called  predictive filter, is combined with
an adaptive constraint margin and a constraint violation cost monitor to minimize the cost of violating soft constraints due to model uncertainty and disturbances. The module can be combined with any controller and is based on minimally modifying the controller output, in a least squares sense, such that constraints are satisfied within the considered horizon. 
The proposed method is computationally efficient and suitable for real-time applications. The effectiveness of
the method is illustrated through a realistic case study of glycosylation constraint satisfaction in continuous monoclonal antibody biosimilar production using  Chinese hamster ovary cells, employing a metabolic network model consisting of 23 extracellular metabolites and 126 reactions. In the case study, the average constraint-violation cost is reduced by more than 60\% compared to the case without the proposed constraint-handling method. 
  \end{abstract}

\end{frontmatter}

\section{Introduction}
Recently, learning-based optimization and control methods have received significant interest \citep{alleyne2023control}, mainly due to their strong learning ability in highly uncertain systems. Such methods have been developed and applied in  applications ranging from robotics to biochemical processes. Some  examples include imitation learning \citep{ravichandar2020recent}, 
learning-based model predictive control \citep{hewing2020learning},  reinforcement learning \citep{ma2021reinforcement,recht2019tour,kim2021model,polydoros2017survey}, Bayesian optimization \citep{shields2021bayesian,mehrian2018maximizing}, iterative learning control and optimization \citep{wang2024iterative,ahn2007iterative}, etc. Learning-based algorithms are in most cases explorative and hence prone to violate constraints that are imposed e.g., for safety or economical reasons.  

In process systems, constraints are typically imposed on variables such as concentrations, temperatures, and pressures, and it is often optimal to operate at or close to such constraints. The constraints can in most cases be temporally violated, but then often with significant economic costs.  
Due to large uncertainties in many process systems, such high-cost constraints are likely to be violated, especially when using explorative learning algorithms.

To minimize the cost of constraint violations in the presence of uncertainty, e.g., disturbances, we propose in this work to combine a predictive filter \citep{wabersich2018linear} with an adaptive constraint margin and a constraint violation cost monitor. 
Predictive filters are used, for example, in autonomous driving applications \citep{ tearle2021predictive}. The margin adaptation employed in this work is based on the margin optimization proposed for economic model predictive control in \cite{TROLLBERG20179065}.  Our main contribution is the combination of the predictive filter and the margin adaptation such that the modular constraint handler can minimize the cost of soft constraint violations by learning the optimal constraint margins directly from real-time data. In order to detect any change in uncertainties that impact on constraint violations, we propose to employ a constraint violation cost monitor and adapt the constraint margin so as to minimize the cost.

A particular strength of the proposed method is that it provides a simple and  universal module that can be directly applied to systems with arbitrary controllers or optimizers. Notably, the proposed method only alters the control signal when necessary, ensuring a minimal modification of the input computed by the controller or optimizer while satisfying constraints, requiring no prior knowledge of the uncertainty. This is of particular importance when employing methods that are explorative, e.g., when learning-based control or optimization methods are applied. The proposed method is computationally efficient and therefore suitable for real-time applications. 

There exist a number of methods to handle constraints in the presence of uncertainty. In standard Model Predictive Control (MPC), e.g., \cite{BorelliBemporadMorar2017}, a fixed nominal model is employed for constrained optimization and uncertainty is dealt with implicitly by adding output feedback. If some prior knowledge on the uncertainty is available, e.g., that it belongs to a given stochastic distribution or some bounded set, stochastic or robust MPC can be employed \citep{bayer2014tube, bayer2016robust,  mesbah2016stochastic, wu2018economic, parisio2016stochastic, lucia2014handling, bayer2018optimal, 6858851}.  When an a priori model of the system dynamics is lacking, model-free methods that give priority to  safety and robustness can be applied, such as safe reinforcement learning \citep{dalal2018safe,garcia2015comprehensive}.  
  A typical case in process control is that a nominal dynamic model is available, but the uncertainty is not well characterized and this causes problems for
 methods that do not adapt to compensate for observed uncertainty during runtime, such as MPC. In such cases one can improve constraint handling by employing learning and adaptive based methods \citep{gahlawat2020l1,bujarbaruah2020adaptive, TROLLBERG20179065, berkenkamp2015safe, oldewurtel2013adaptively, chachuat2008process}. However, essentially all such methods
rely on particular controller structures, such as different variants of MPC, and are therefore not applicable to other control algorithms. Predictive filters \citep{ tearle2021predictive, wabersich2018linear}, on the other hand, are modular and can be combined with essentially any control algorithm but are focused on satisfying hard safety constraints. 
Predictive filters have  recently been considered also for dealing with soft constraints, but then primarily to avoid infeasibilities \citep{wabersich2021predictive,9449855}.   There also exist works on robust extensions of the predictive safety filter, e.g., \cite{wabersich2021predictive_auto}, but these are all based on prior knowledge of the uncertainty. We note that  constraints also can be handled with other methods, such as control barrier functions \citep{ames2019control} and Hamilton–Jacobi reachability analysis \citep{fisac2018general}.   The interested reader is referred to \cite{wabersich2023data} and \cite{brunke2022safe} for an overview of constraint handling methods.  Finally, in the context of learning-based control, there exist works focusing on parameter adaptation with economic considerations \citep{kim2021model,kordabad2021reinforcement, gros2019data, kordabad2021mpc, alhazmi2021reinforcement, zanon2020safe, piga2019performance}. However, rather than high-cost constraint satisfaction for arbitrary systems, they consider obtaining operational conditions to optimize economical performance using learning based methods. To our knowledge there exist no methods for constraint satisfaction  that can be applied to systems with arbitrary controllers while adapting optimal constraint margins for the soft constraints directly from data, requiring no prior knowledge on the uncertainty.  

We stress that the proposed method is not intended to replace MPC, but rather to serve as a modular constraint-handling layer that can be added on top of any controller. This is especially valuable in settings where constraints must be handled even though the underlying controller—such as legacy industrial logic or modern learning-based strategies—cannot handle constraints explicitly. 

Preliminary results have been presented in \cite{wang2022modular} using  examples involving   water tank temperature control  and  fed-batch penicillin fermentation optimization. Here we extend these results and present a realistic case study of glycosylation constraint satisfaction in continuous monoclonal antibody biosimilar production based on a realistic metabolic network and glycosylation models.  

The rest of the paper is organized as follows. We first   provide an introductory example of  glycosylation constraint satisfaction in Section~\ref{sec:motivating_example}. We  formulate the considered problem in Section~\ref{sec:problem_formulation}, and introduce the proposed constraint handler  in Section~\ref{sec:adaptive_predictive_filter}. In Section~\ref{sec:example}, we illustrate the effectiveness by revisiting the introductory example. Finally, we provide conclusions and discussion of the work in Section~\ref{sec:conclusion}.

\section{Introductory example}
\label{sec:motivating_example}
We present an introductory example to illustrate the impact of uncertainty on constraint handling and the need for constraint
handling methods that can learn optimal constraint margins from real-time data and, furthermore, be used in combination with arbitrary controllers.
 
Glycosylation is a common method to modify the properties of proteins by adding carbohydrate components to form glycoproteins. We
here consider glycosylation in continuous therapeutic monoclonal antibody (mAb) biosimilar production operating around steady-state conditions and in particular
constraints on the level of $G0$ glycoforms. Regulatory agencies impose a required range that the $G0$ level should be within for
safety and efficacy reasons \citep{christl2017biosimilars,kunert2016advances,EUMedAg2014}.  
Assume process optimization gives an optimal $G0$ percentage level $P_{G0,opt} = 35\%$, which coincides with the lower limit of the $G0$ percentage. If the $G0$ percentage  $P_{G0}$ is less than the optimal value $P_{G0,opt}$,  the product is substandard and must be rejected, resulting in a significant economic loss. To track $P_{G0,opt}$ in the presence of disturbances, we apply a simple feedback control strategy to adjust glucose concentration in the feeding medium based on real-time estimates of steady-state $P_{G0}$ obtained from in-line measurements of glucose consumption rates. More details will be provided when we return to this process towards the end of the paper.  

\begin{figure}
    \includegraphics[width = 8.4cm]{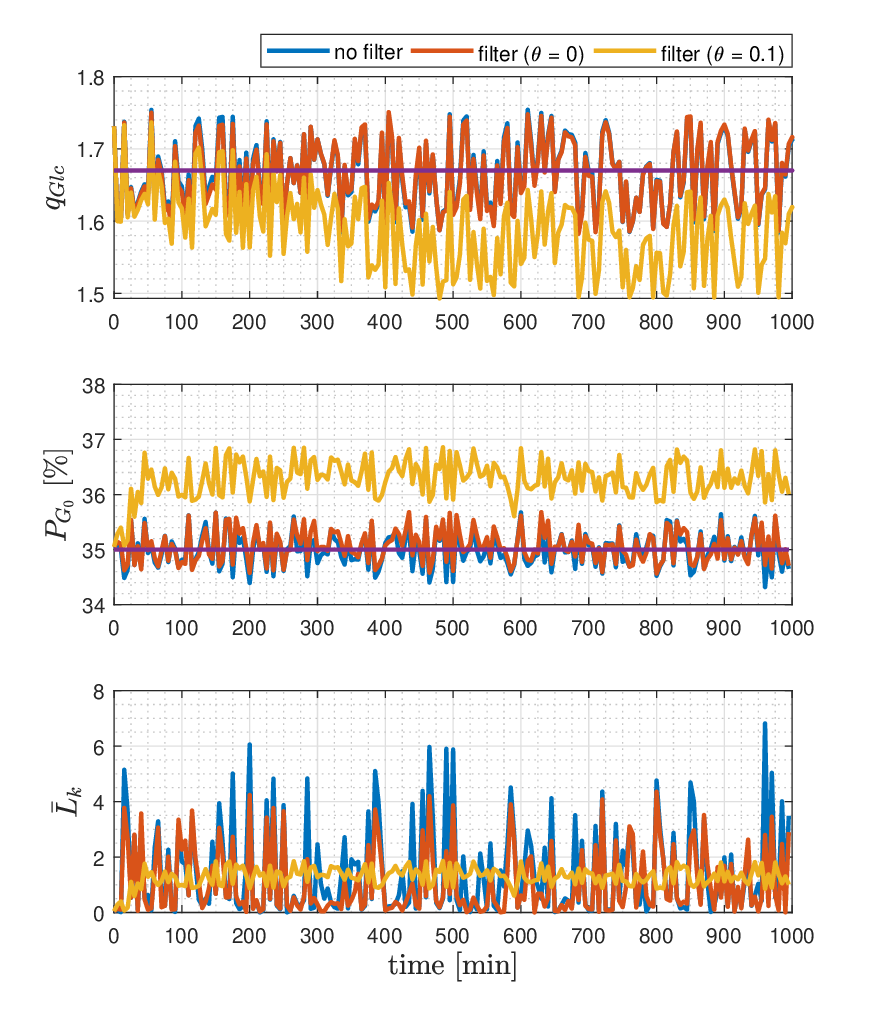} 
\caption{The constraint violation of $q_{Glc}$ (top),  $P_{G_0}$ performance (middle) and corresponding economic loss $\bar{L}_k$ (bottom, described later) with feedback controller (blue),  feedback controller with predictive filter with $\theta = 0$ (red) and  $\theta = 0.1$ (yellow) in the presence of uncertainties including disturbances. The average economic losses  over the considered period are 1.43, 1.06  and 1.28, respectively.  The purple lines represent $q_{Glc,ref} = 1.67$ $\text{pmol/(cell$\cdot$day)}$ (top figure) and  the corresponding optimal value $P_{G_0,opt} = 35\%$ (middle figure). The considered economic loss  includes both the cost of constraint violation and the cost of operating away from $P_{G_0,opt} = 35\%$.}  
\label{fig:introductory_example}
\end{figure}

As shown in Figure~\ref{fig:introductory_example} (in blue), when applying the feedback controller to track $P_{G0,opt}$, there are, as expected, frequent violations of the constraint $P_{G0} > P_{G0,opt}$, and are reflected in a higher economic loss in the lower part of the figure\footnote{Note that we here consider constraints on the instantaneous production in a continuous process, rather than on the batch average as is often considered in biopharmaceutical batch production.}.  The main reason for the repeated constraint violations is the presence of various disturbances.

In order to handle the constraints, we add a separate control module named predictive filter \citep{wabersich2018linear} which modifies the control signal generated by the feedback controller if a constraint violation is predicted within the considered horizon. The prediction requires a dynamic model that can compute future values of the $P_{G0}$ given the current state of the system. For this purpose, we here employ the model established in \cite{zhang2021control,zhang2020glycan} and \cite{hagrot2019novel}, and the constraint is imposed on the glucose consumption rate $q_{Glc}$ which is assumed measured and inversely related to $P_{G0}$ \citep{zhang2021control}.   
As can be seen from  Figure~\ref{fig:introductory_example} (in red), the constraint violations are now somewhat less severe but still frequent, mainly due to the fact that future disturbances cannot be predicted combined with model uncertainty, i.e., discrepancies between the model used for predictions and the true process.

Since constraint violations are inevitable in the presence of unmeasured disturbances and model uncertainty, it is advisable to introduce a constraint margin, i.e., to force nominal operation some distance away from the constraint. 
For this purpose, we introduce a constraint margin  $\theta \in \mathbb{R}$ such that the modified constraint in the predictive filter is 
$q_{Glc} \leq q_{Glc,ref}-\theta$.
As shown in Figure~\ref{fig:introductory_example} (middle, in yellow), with $\theta = 0.1$, there is now no violation of $P_{G_0,opt}$. However, the constraint margin is conservative since the resulting $P_{G_0}$ is far from the optimum $P_{G_0,opt}$, which also results in economic losses.  Thus, it is important to adapt the margin such that the combined cost of constraint violations and cost of operating away from the optimum is minimized.

 \section{Problem Formulation}
 \label{sec:problem_formulation}
Consider a discrete-time nonlinear process model
\begin{equation}\label{eq.sys}
    {x}_{k+1} = f(x_k,u_k) + \Delta_k, 
\end{equation}
where $k \in \mathbb{N}$ represents the time step of the discrete-time system; $x \in \mathbb{R}^{n_x}$ is the state vector;  $u \in \mathbb{R}^{n_u}$ is the control input; $f(\cdot): \mathbb{R}^{n_u} \times  \mathbb{R}^{n_x} \to \mathbb{R}^{n_x}$ represents the nonlinear dynamics of the process; $\Delta_k \in \mathbb{R}^{n_x}$ represents the uncertainty  at time step $k$. A key assumption of this paper is that  the magnitude and characteristics of $\Delta_k$ is unknown a priori. 
The system is subject to constraints 
\begin{equation}
\begin{split}
        g_h(x_k,u_k) & \leq 0,   h(x_k,u_k) = 0, \\ g_e(x_k,u_k) & \leq 0,
\end{split}
\end{equation}
where $ g_h(x_k,u_k) \leq 0$ and $h(x_k,u_k) = 0$ are  hard inequality and equality constraints enforced by physical limitations, while $g_e(x_k,u_k) \leq 0$ are imposed inequality constraints that can be temporarily violated but  then with a significant violation cost.  Let the cost of violation of $g_e(x_k,u_k) \leq 0$ at sample $k$ be
\begin{equation}\label{eq.violate_cost_measure}
   L_k = \lambda^T \max(0,g_e(x_k,u_k)),
\end{equation}
where $\lambda \in \mathbb{R}^{n_{\lambda}}$ is a constant vector that reflects the constraint violation cost.

We here assume that the system \eqref{eq.sys} is regulated by a, possibly learning-based,  controller that generates a desired  control signal of interest $u^d_k$ at sample $k$. Although constraints can be integrated into the controller computations using e.g., MPC, we consider the case where the controller does not handle  constraints, since we aim at solving constraint handling with general control algorithms. Rather, we consider adding a separate algorithm that
predicts future states and modifies the computed $u_k^d$ in case of a predicted violation. The modification of $u^d_k$ should be kept at a minimum subject to satisfying the constraints.

Ideally, in the absence of disturbances and model uncertainty, the modification of the control input should be such that future constraints are just satisfied as this will correspond to the minimal modification of the computed input. However, the presence of disturbances and model uncertainty implies that constraints are likely to be violated despite modifying the input such that there are no predicted violations. Furthermore, since the cost of violations can be highly significant, it will in most cases be optimal to operate some distance away from the constraints. The aim here is therefore to determine the modification of $u^d_k$ such that the long-term cost of violating $g_e(x_k,u_k) \leq 0$, in the presence of disturbances and model uncertainty, is minimized. The cost is defined as
\begin{equation}\label{eq.violate_cost_true}
{J} = \lim_{n\to \infty} \frac{1}{n} \sum_{k = 1}^{n} \bar{L}_k,
\end{equation}
with 
\begin{equation}
\label{eq.violate_cost_measure_with_penalty}
    \bar{L}_k = L_k + P_k,
\end{equation}
where $L_k$ is the cost of violating the constraint, as defined in \eqref{eq.violate_cost_measure}, and $P_k$ reflects the cost of operating away from the optimal operational conditions at sample $k$. The purpose of $P_k$  is to avoid over-conservative constraint margin and make a trade-off between constraint satisfaction and optimal operating conditions.

Since we aim at solving constraint violation problems in the general case we do not assume linear dynamics, independent and identically distributed disturbance, etc. Rather we make the following reasonable assumptions
\begin{assum}
\label{assumption}
We assume that 
\begin{enumerate}
    \item the size and characteristics of $\Delta_k$ is unknown a priori;
    \item the closed-loop system is stable;
    \item there exists at least a local optimum of the cost \eqref{eq.violate_cost_true}.
\end{enumerate}
\end{assum}
In this work, we  propose a method  based on the predictive filter \citep{wabersich2018linear} combined with online learning of constraint margins and a violation cost monitor to solve the problem, under Assumption~\ref{assumption}. 

\section{Learning-based predictive filter}
\label{sec:adaptive_predictive_filter}
In the absence of uncertainty, the constraint satisfaction problem can be solved using a safety predictive filter \citep{ tearle2021predictive, wabersich2018linear}. In the presence of uncertainty, one can employ robust extensions of the safety predictive filter provided a priori knowledge on the uncertainty magnitude and characteristics are available \citep{wabersich2021predictive_auto}. However, such a priori knowledge on the uncertainty is usually not available in practice, which leads to challenges when implementing robust constraint handling methods \citep{brunke2022safe}. Robust methods also tend to be overly conservative. 

\begin{figure}[htbp]
    \centering
    \includegraphics{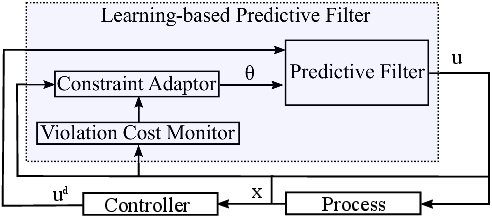}
    \caption{The learning-based predictive filter framework, where $x$ denotes the measured or estimated states.}
    \label{fig:adaptive_predictive_filter_framework}
\end{figure}

We here propose a learning-based  predictive filter, as illustrated in Figure~\ref{fig:adaptive_predictive_filter_framework}.   
Basically, we introduce a constraint margin $\theta \in \mathbb{R}^{n_{\lambda}}$ for the high-cost constraints, i.e.,  $g_e(x_k,u_k) +\theta \leq 0$, which is learned online directly from available data using the constraint adaptor seeking the margin $\theta$ that minimizes the cost \eqref{eq.violate_cost_true}. Given a learned $\theta$, the predictive filter then computes the minimum modification of the control input to satisfy the constraint with margin $\theta$. After $\theta$ has converged to an optimal value using the constraint adaptor, a violation cost monitor constantly monitors the high-cost constraint violation to determine whether $\theta$ should be re-learned based on changes in the uncertainty size and characteristics.  

We next introduce the different parts of the proposed learning based predictive filter: the predictive filter, the constraint adaptor and the constraint violation cost monitor.

\subsection{Predictive filter}
\label{sec:predicitve_filter}
Given a  constraint margin $\theta$, we apply a predictive filter to improve constraint satisfaction $g_e(y_k,u_k)  +\theta \leq 0$ by modifying the control signals generated by the controller so as to avoid predicted constraint violations. The filter works in a receding horizon fashion, and a new prediction is made at every sample instance. 

The predictive filter \citep{wabersich2018linear} is completely dedicated to ensure constraint satisfaction. The basic idea is that, given a desired control signal and state estimate at the current sample, the filter first determines whether it is possible to find a control sequence, starting with the desired control signal at the current sample, that will satisfy all constraints within the given horizon. If  all constraints can be satisfied within the horizon, the desired control signal is applied directly. On the other hand, if a constraint violation is predicted to be unavoidable with the computed current control input, the closest current control input in a least-square sense that allows for a control sequence that satisfies all constraints over the control horizon is determined. Formally, given a constraint margin $\theta$, the predictive filter is given by
\begin{equation}\label{eq.predict_filter}
\begin{aligned}
\min_{u_{i|k}} \quad & \|u^d_k - u_{0|k} \|_2 + V_p( \epsilon_{0|k}, \epsilon_{1|k}, \dots, \epsilon_{N|k})\\
\textrm{s.t.} \quad & \text{for all } i = 0,\dots, N:\\
  & x_{0|k} = x_k,\\
  & x_{i+1|k} = f(x_{i|k},u_{i|k}),\\
  &  g_h(x_{i|k},u_{i|k})   \leq 0,   h(x_{i|k},u_{i|k})  = 0, \\
  & g_e(x_{i|k},u_{i|k})  +\theta  \leq \epsilon_{i|k},  \epsilon_{i|k} \geq 0
\end{aligned}
\end{equation}
where $u^d_k$ is the desired control at the current time step $k$ generated by the controller; $N$ is the prediction horizon of the predictive filter; the subscript $i|k$ represents the $i^{th}$ step ahead prediction with the predictive filter initialized at time step $k$;  $V_p( \epsilon_{0|k}, \epsilon_{1|k}, \dots, \epsilon_{N|k})$ represents the penalty function on slack variables  $\epsilon_{i|k}  \in \mathbb{R}^{n_{\lambda}}$, whose elements are non-negative and represent slack variables on the soft inequality constraints to solve potential infeasibility issues. The predictive filter generates the optimal control sequence $\{u_{i|k}^*\}$ of which the first element $u_{0|k}^*$ is applied to the process.  The filter is re-applied at the next time step in a receding horizon fashion. 

We do not provide any stability proof for the predictive filter here, but simply assume that  the modification of the desired control input by the predictive filter is bounded in a way such that the closed-loop system remains stable for any feasible choice of $\theta$. We refer the interested readers to e.g., \cite{wabersich2018linear} and \cite{wabersich2021predictive_auto} for discussions on stability and feasibility of predictive filters. 

With the predictive filter in place, the remaining problem is then  to determine the optimal constraint margin $\theta$ that minimizes the impact of uncertainties on constraint violations without being overly conservative. For this purpose we propose a constraint adaptor to learn the optimal constraint margin, largely based on ideas presented in \cite{TROLLBERG20179065}.

\subsection{Constraint adaptor}\label{sec.constraint_adaptor}
The purpose of the constraint adaptor is to learn the optimal constraint margin $\theta$ that minimizes \eqref{eq.violate_cost_true} based on online measurements. To obtain gradient information from available, possibly noisy and transient, data, we propose to employ finite-difference stochastic approximation (FDSA) \citep{spall1998implementation,spall2005introduction,kushner2012stochastic}.  
Below we briefly outline the FDSA-based adaptation method. We refer interested readers to \cite{TROLLBERG20179065} for further details.

We are seeking a local, possibly global, minimum of the objective function $J$ defined in \eqref{eq.violate_cost_true} with respect to the constraint
margin $\theta$. We assume that $J$ is well defined  for any feasible choice of $\theta$ and want to determine the local gradient $\p J/\p \theta$. Since we do not have a closed form expression for the dependency of $J$ on $\theta$ available, we seek to determine the gradient from measurement data. The main challenges with approximating the gradient based on measurement data is partly the presence of measurement noise in the data and partly that the obtained data typically are obtained under transient conditions leading to a bias in the gradient estimate. The FDSA method can handle both these issues in an efficient manner.

At the $j^{th}$ iteration of the FDSA update, the $l^{th}$ component of the finite-difference approximation of the gradient $\p J/\p \theta$ is given by
{\begin{equation}
    [DJ(\theta_j,c_j)]_l = \frac{{J}_{j,l}^+ - {J}_{j,l}^-}{2c_j},
\end{equation}
where $\theta_j$ is the constraint margin value at the $j^{th}$ iteration; $c_j$ is the perturbation size on $\theta$ to approximate the gradient using finite differences;   ${J}_{j,l}^+$ and ${J}_{j,l}^-$ are the long-term constraint violation cost \eqref{eq.violate_cost_true} with $\theta_j + c_j e^l$ and $\theta_j - c_j e^l$ as constraint margins in the predictive filter, respectively,  $e^j$ being the standard basis vector. The constraint violation 
costs can be approximated by averaging online measurements of \eqref{eq.violate_cost_measure_with_penalty} over large samples.}

Similar to gradient descent,  the FDSA based method recursively updates $\theta$ based on the gradient estimate
\begin{equation}
\label{eq.adapt_theta}
    \theta_{j+1} = \theta_j + a_j DJ(\theta_j,c_j),
\end{equation}
where $a_j$ is the step size along the gradient approximation $ DJ(\theta_j,c_j)$. The convergence of $\theta$ can be guaranteed  under certain conditions on the step size sequences $\{a_j\}$ and $\{c_j\}$ \citep{spall1998implementation,spall2005introduction}: 
\begin{enumerate}
    \item $c_j \to 0$ as $j \to \infty$, to diminish the error in the finite-difference approximation of the gradient over time;
    \item $\lim_{n\to\infty} \sum_{j = 1}^n a_j = \infty$, to avoid premature convergence;
    \item $\lim_{n\to\infty} \sum_{j = 1}^n a_jc_j < \infty$ and  $\lim_{n\to\infty} \sum_{j = 1}^n a_j^2/c_j^2 < \infty$, to reduce the adaptation rate by ensuring the step sizes decrease over time. 
\end{enumerate}
The parameters in FDSA can be selected following standard guidelines \citep{spall1998implementation,spall2005introduction}.  \citet{TROLLBERG20179065} also discuss practical heuristics for parameter tuning in FDSA implementations. Also note that the FDSA-based method uses a data-based finite-difference approximation that estimates the gradient directly from measurement based evaluations of $J$ and therefore does not require any explicit derivatives of $J$. 
If \begin{equation}
\label{eq:stop_criteria}
    |\theta_{j+1}-\theta_j| < \rho_{learning},
\end{equation} where $\rho_{learning}$ is a user-defined threshold value, it implies that the constraint margin has reached an optimal value based on available information about uncertainty. In such a case,  the constraint adaptation of $\theta$ will stop until the constraint violation cost monitor re-initiates the learning process, as we will introduce next. {Due to the time-consuming and expensive nature of biopharmaceutical experiments—and because stable experimental conditions are required for large-scale production—we stop updating $\theta$ once \eqref{eq:stop_criteria} is satisfied. In applications where sample collection is less costly, it is of course feasible to continue adapting $\theta$. We note that stochastic variations may lead to satisfaction of \eqref{eq:stop_criteria} at a random instance. A more robust alternative stopping criterion could be, for example, to require that condition \eqref{eq:stop_criteria} be satisfied for several consecutive samples.}  

As we will demonstrate below, the FDSA-based gradient descent can converge relatively fast to the optimal $\theta$, and hence the optimal margin can be learned  efficiently in online applications. Again, we do not provide any stability proof here, but note that the FDSA method will slow down over time which implies that a separation of time scales will appear over time and this will improve stability \citep{TROLLBERG20179065}.

Note that there exist alternatives to learning optimal constraint margins from data in order to minimize the impact of high-cost constraint violations as considered in this work. These include methods based on updating dynamic models and introducing correction terms in the optimization, see e.g., \citep{marchetti2016modifier,darby2011rto} for an overview. However, all these methods rely on particular controller structures and focus on obtaining the operational conditions that optimize the overall economical performance, rather than focusing on high-cost constraint satisfaction with a minimal modification of the desired inputs generated by arbitrary controllers, such as learning-based controllers, which is the focus of this work.

\subsection{Constraint violation cost monitor}
\label{sec:cost_monitor}
When $\theta$ has converged to an optimal value using the constraint adaptor, we apply the converged constraint margin $\theta^*$ in the predictive  filter. Simultaneously, the constraint violation cost should be monitored constantly in order to detect changes in the size and characteristics of disturbances and uncertainty and re-initiate adaptation of $\theta$ if needed.

For this purpose, we first obtain the average constraint violation cost for the converged optimal constraint margin $\theta^*$, by  averaging the constraint violation cost $\bar{L}_k$ in \eqref{eq.violate_cost_measure_with_penalty} over an horizon corresponding to $n_{monitor}$ samples since $\theta^*$  were first applied
\begin{equation}
\label{eq:average_constraint_violation}
    J^* =   \frac{1}{n_{monitor}} \sum_{k = 1}^{n_{monitor}} \bar{L}_k.
\end{equation}
Subsequently, we proceed to monitor the average violation cost by obtaining  the 
monitored average constraint violation cost when a new set of $n_{monitor}$ samples is collected 
\begin{equation}
\label{eq:average_constraint_violation_monitor}
    J_{monitor} =   \frac{1}{n_{monitor}} \sum_{k = 1}^{n_{monitor}} \bar{L}_k.
\end{equation}
 If \begin{equation}
 \label{eq:monitoring_condition}
     |{J_{monitor} - J^*}| > \rho_{monitor}|{J^*}|, 
 \end{equation} where $\rho_{monitor} \in (0,1)$ is a user-defined threshold, it indicates that  the current constraint margin $\theta^*$ may no longer be optimal due to changes in the uncertainties, and hence a new optimal $\theta$ should be learnt using the constraint adaptor.  

\subsection{Summary of learning-based predictive filter}
The proposed learning-based predictive filter framework can be summarized as follows.
\begin{enumerate}
    \item select an initial constraint margin $\theta_0$; prediction horizon $N$ and sampling time $T_s$ in the predictive filter;  step size sequences $\{a_j\}$ and $\{c_j\}$ and threshold for convergence $\rho_{learning}$ in the constraint adaptor; the number of  samples for averaging $n_{monitor}$ and the threshold to detect the change of uncertainties $\rho_{monitor}$ in the constraint violation monitor;
    \item learn the optimal constraint margin $\theta^*$ using the constraint adaptation described in Section~\ref{sec.constraint_adaptor}, until $|\theta_{j+1}-\theta_j| < \rho_{learning}$; 
    \item apply the converged $\theta^*$ as the optimal constraint margin in the predictive filter described in Section~\ref{sec:predicitve_filter}, and calculate  the average constraint violation cost $J^{*}$ according to  \eqref{eq:average_constraint_violation};
    \item monitor the constraint violation cost using  the  monitor described in Section~\ref{sec:cost_monitor}: 
    \begin{enumerate}
        \item calculate the monitored average constraint violation cost $J_{monitor}$ according to  \eqref{eq:average_constraint_violation_monitor};
        \item check condition \eqref{eq:monitoring_condition}
        \begin{itemize}
             \item if $|{J_{monitor} - J^*}| > \rho_{monitor}|{J^*}|$, 
        re-learn optimal constraint margin by repeating Step 2-3;
    \item otherwise, keep  $\theta^*$  as the constraint margin applied in the predictive filter and repeat Step 4 when a new set of $n_{monitor}$ samples is collected. 
        \end{itemize}
       
    \end{enumerate}
\end{enumerate}

\subsection{Guideline for parameter selection}
The parameters of the proposed method can be selected as follows.
The initial constraint margin $\theta_0$ is typically set to $\theta_0 = 0$, after which the constraint adaptor learns the optimal margin from online data.
The prediction horizon $N$ and sampling time $T_s$ in the predictive filter are chosen based on the available sampling rate and computational capacity, with larger $N$ improving prediction accuracy but increasing computational load. The step-size sequences $\{a_j\}$ and $\{c_j\}$ follow standard FDSA guidelines \citep{spall1998implementation,spall2005introduction} and may be tuned heuristically \citep{TROLLBERG20179065} if needed.
The convergence threshold $\rho_{\text{learning}}$ determines when the margin is considered converged, where smaller values lead to slower but more precise convergence. The averaging window $n_{\text{monitor}}$ should be sufficiently large to reduce stochastic variation in the violation cost, while the detection threshold $\rho_{\text{monitor}}$ specifies when the margin should be re-learned.
Smaller $\rho_{\text{monitor}}$ values increase sensitivity to changes in the underlying uncertainty.  

\section{Case Study: Glycosylation in Perfusion BioReactor}
\label{sec:example}

To demonstrate the effectiveness of the proposed method, we revisit  the introductory example in Section~\ref{sec:motivating_example}. 
\subsection{Glycosylation in Continuous mAb Production}
Glycosylation, a post-translational modification of proteins produced by mammalian cells, is a critical quality attribute (CQA) of monoclonal antibodies (mAbs). It has significant impact on properties such as the pharmacokinetics and pharmacodynamics of mAbs  \citep{sha2016n,liu2015antibody}. The glycosylation attributes of  biosimilar products are required to be within given reference ranges --- so-called specifications --- to ensure product safety and efficacy \citep{christl2017biosimilars,kunert2016advances,EUMedAg2014}. If the reference ranges are exceeded, the biosimilar products needs to be rejected or recycled, or it must be justified that the biosimilar has a similar effect as the reference product for approval. In any case, violating the given reference range will induce significant economic losses. 

Controlling glycosylation such that it is maintained within a given reference range is a challenging task. This is particularly the case in the production of biopharmaceuticals, where product quality can be affected by medium components or physical conditions when the reactor operates close to one of the reference limits—a situation that is generally economically optimal provided specification violations can be avoided.  
Although many  methods  have been proposed  for controlling glycosylation \citep{zhang2021control,kappatou2020quality,kotidis2019model}, most methods are based on open-loop control and hence sensitive to disturbances and model uncertainties.  Recent advances in  monitoring tools, including in-line Raman spectroscopy \citep{silge2022trends,schwarz2022monitoring,domjan2022real,tharmalingam2015framework,pais2014towards},  and the development of mathematical models of glycosylation  \citep{vstor2021towards,luo2021bioprocess,zhang2020glycan}, have facilitated the application of closed-loop feedback control methods, such as MPC \citep{zupke2015real}. However, most control methods considered in bioprocessing do not explicitly consider constraint handling, neither do  emerging explorative learning-based methods, and hence there is a need to add such a feature to the control system to avoid significant costs
imposed by violating the imposed regulatory reference limits \citep{wang2024iterative,ma2021reinforcement,kim2021model,mehrian2018maximizing}. 

We here formulate the glycosylation reference limits as soft constraints 
that  can be temporarily violated but then with an economic cost that is penalized when determining the optimal constraint margin to operate at. 
The particular problem we consider is maintaining the level of the glycoform $G0$, reflecting the degree of glycosylation maturation \citep{zhang2020glycan,sha2016n}, within a given reference range during continuous mAb production, which enables higher productivity over extended operating periods compared with traditional batch processes \citep{chotteau2015perfusion}.  
Note that product specifications on biopharmaceuticals typically are defined based on a batch averaged quality, but that the introduction of continuous perfusion processes makes it feasible to impose quality specifications on the instantaneous production. 
In this case study we consider the latter case. We assume that in-line measurements of glucose consumption rates are available and that the level of glycosylation can be estimated based on these measurements and the model presented in \cite{zhang2021control,zhang2020glycan}.

\subsection{Modeling}
\label{sec:metabolic_network_model}
\begin{figure}
    \begin{center}
    \includegraphics[width = 8.4 cm]{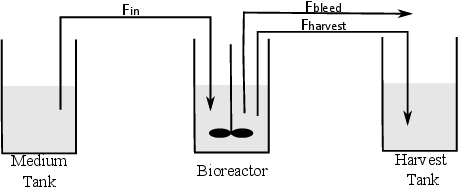}    
    \caption{Schematic of a perfusion process.}  
    \label{fig:perfusion}
    \end{center}
\end{figure}

\begin{figure}
    \begin{center}
    \includegraphics[width = 8.4 cm]{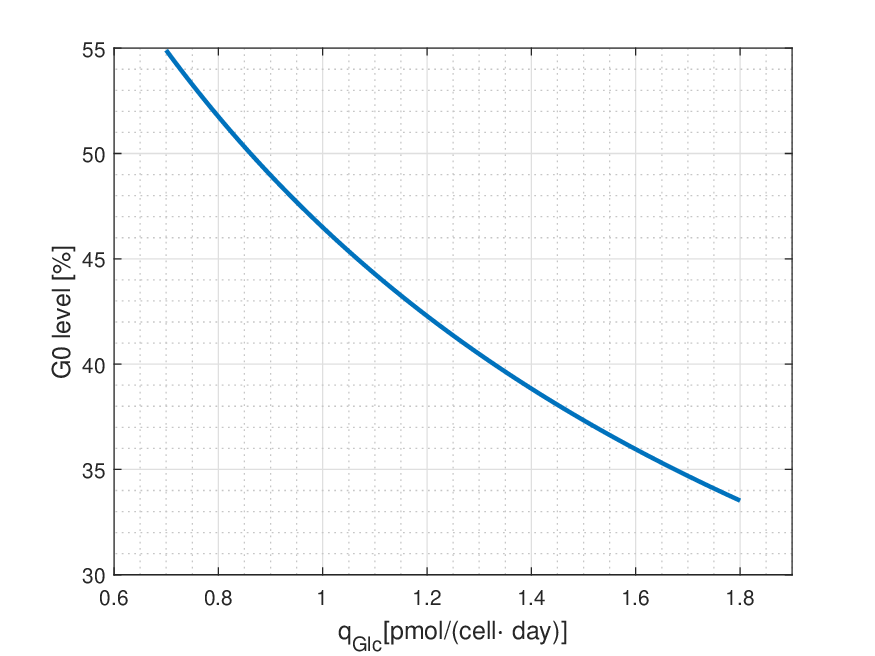}    
    \caption{G0 level glycoform percentage $P_{G_0}$ as a function of glucose consumption rate $q_{Glc}$.}  
    \label{fig:Glc_G0_level}
    \end{center}
\end{figure}

We consider glycosylation constraint satisfaction problem in continuous bioproduction, known as perfusion operation,  as illustrated in Figure~\ref{fig:perfusion} \citep{chotteau2015perfusion,bielser2018perfusion}. The mass balance equations  \citep{karst2017modulation} are given by
\begin{equation}
\label{eq:mass_balance_mab}
 \small
\begin{split}
\frac{d\text{V}[\text{M}_\text{i}]}{d\text{t}} & = \text{q}_{\text{M}_\text{i}}\text{V}[\text{X}_\text{v}] + \text{F}_\text{{in}}[\text{M}_\text{{i,in}}] - \text{F}_\text{{out}}[\text{M}_\text{i}], \\
\frac{d\text{V}[\text{mAb}]}{d\text{t}} & = \text{q}_\text{{mAb}}\text{V}[\text{X}_\text{v}] - \text{F}_\text{{out}}[\text{mAb}],\\
\frac{d\text{V}[\text{X}_\text{v}]}{d\text{t}} & = \text{q}_\text{{biomass}}\text{V}[\text{X}_\text{v}] - \text{F}_\text{{bleed}}[\text{X}_\text{v}],
\end{split}
\end{equation}
with $\text{F}_\text{{out}}  =   \text{F}_\text{{bleed}} + \text{F}_\text{{harvest}}$. 
The unit and meaning of all variables and parameters are summarized in Table~\ref{tab:parameters_mass_balance_table}. 
To determine the production and consumption rates of the various external metabolites, we consider  the metabolic network model developed in  \cite{hagrot2019novel} based on experimental data presented in \cite{hagrot2017poly}. The model involves 23 extracellular metabolites and 126  reactions   in mAb-producing mammalian CHO  cells. Under a pseudo steady-state assumption \citep{quek2010metabolic},  the external metabolite rates are given by
\begin{equation}
\label{eq:rate_equation}
 q_{ext} = A_{ext} Ew,
\end{equation}
where  $q_{ext} = \begin{bmatrix}  q_{M_1}, \cdots,  {q_{M_{21}}}, q_{biomass}, q_{mAb} \end{bmatrix}^\top$; $ A_{ext} \in \mathbb{R}^{23\times 126}$ is  the stoichiometric matrix; $E \in \mathbb{R}^{126\times 132}$ is a matrix with  elementary flux modes related to the macro-reactions \citep{schuster1994elementary}; $w \in \mathbb{R}^{132}$ is a  vector whose  $l^{th}$ element, $w_l$, is the macroscopic flux over macro-reaction $l$  modeled as a product of  Monod  functions
\begin{equation}
\label{eq:monod_equation}
\begin{split}
\small{w_l  =   w_{max,l} \underbrace{\prod_{i\in M_{ext,s,l}} \frac{ \frac{[M_i]}{K_{s,i,l}}}{ \frac{[M_i]}{K_{s,i,l}}+1}}_\text{substrate saturation}   \underbrace{\prod_{i\in M_{ext,p,l}} \frac{1}{ \frac{[M_i]}{K_{p,i,l}}+1}}_\text{product inhibition} \underbrace{\prod_{i\in M_{ext,r,l}} \frac{1}{ \frac{[M_i]}{K_{r,i,l}}+1}}_\text{metabolite inhibition}.}
\end{split}
\end{equation}
Having obtained the consumption rates from \eqref{eq:mass_balance_mab} and \eqref{eq:rate_equation}, 
the steady-state G0-level glycoform percentage $P_{G0}$ --- representing the proportion of G0 and G0F among all fucosylated and afucosylated glycans --- is estimated as a function of the glucose consumption rate $q_{\mathrm{Glc}}$ 
using a steady-state model from \citet{zhang2021control}, assuming glucose is the only fed carbon source:
\begin{equation}
P_{G0}(q_{\mathrm{Glc}})=
\left[
1 -
\frac{
P_{G1F}(q_{\mathrm{Glc}})+
P_{G2F}(q_{\mathrm{Glc}})+
P_{G1}(q_{\mathrm{Glc}})}
{P_{\mathrm{nonHM}}(q_{\mathrm{Glc}})}
\right]\times100\%,
\label{eq:PG0_main}
\end{equation}
where the galactosylated glycoform fractions $P_{G1F}$, $P_{G2F}$, and $P_{G1}$ follow Michaelis--Menten-type functions of  $q_{\mathrm{Glc}}$. The non–high-mannose fraction is given by
\begin{equation}
P_{\mathrm{nonHM}}(q_{\mathrm{Glc}})
=
P_{\max}\,
\Phi_{\mathrm{nac}}(q_{\mathrm{Glc}})\,
\Psi_{\mathrm{man}}(q_{\mathrm{Glc}}),
\label{eq:PnonHM}
\end{equation}
with $\Phi_{\mathrm{nac}}(\cdot)$ and $\Psi_{\mathrm{man}}(\cdot)$ 
representing saturation and inhibition effects associated with N-acetylglucosamine (GlcNAc) and mannose precursors respectively, 
each described by Michaelis--Menten-type kinetics. 
$P_{\max}$ represents the upper limit of non-high-mannose glycans. Parameter values are adopted from \citet{zhang2020glycan}. The static relationship is illustrated in Figure~\ref{fig:Glc_G0_level}. It
serves as a static proxy that relates the steady-state glycosylation quality to glucose consumption. 
The interested reader is referred to \cite{zhang2021control,zhang2020glycan} and \cite{hagrot2019novel}  for details on the considered model.  

In the predictive filter, we assume that (i) in-line measurements of consumption rates are available at a 5-min sampling interval (e.g., via Raman spectroscopy), and (ii) the steady-state glycosylation pattern can be  estimated from these measurements using the static model in \citet{zhang2021control,zhang2020glycan}. This static model has been experimentally validated in continuous mAb production \citep{zhang2021control}, making such measurements realistic and implementable in practice. If the glycosylation model is inaccurate, data from previous experiments could be used to construct confidence intervals or quantify estimation uncertainty, enabling a robust extension of the proposed method. Addressing such extensions, however, lies beyond the scope of the present study.

\begin{table}
 \small
\begin{center}
\caption{Unit and description of parameters and constants in   \eqref{eq:mass_balance_mab}. } \label{tab:parameters_mass_balance_table}
    \begin{tabular}{p{1.5cm}p{1.7cm}p{4.2cm}}
    \hline
  Constant & Unit & 
Description \\
    \hline
    $V = 1$ & $\text{RV}$    & bioreactor volume \\
        $\text{F}_\text{{in}} = 1$ & $\text{RV/day}$     &feed flow rate into bioreactor\\
         $ \text{F}_\text{{out}} = 1$ & $\text{RV/day}$    & flow out of the bioreactor \\
         $[\text{X}_\text{v}] = 100$ & $\text{mM}^*$  & viable cell density in bioreactor\\
             \hline
    Parameter & Unit & 
Description \\
    \hline
          $ \text{F}_\text{{bleed}}$& $\text{RV/day}$    & bleed rate  to maintain desired viable cell density\\
          $\text{F}_\text{{harvest}}$ & $\text{RV/day}$   & harvest rate  to collect product (mAb) \\          
           $[\text{M}_\text{{i,in}}]$ & $\text{mM}$ &   concentration of extracellular metabolite $M_i$  in  feed flow cultivation medium\\
          $[\text{M}_\text{i}]$ & $\text{mM}$ &  concentration of extracellular metabolite  $M_i$ excluding biomass and mAb in bioreactor\\   
          $[\text{mAb}]$ & $\text{mM}$ &  mAb concentration in  bioreactor\\
          $\text{q}_{\text{M}_\text{i}}$  & $\text{pmol/(cell$\cdot$day)}$ &  uptake (or secretion)  rate of extracellular metabolite  $M_i$ \\
           $\text{q}_\text{{biomass}}$  & $\text{pmol/(cell$\cdot$day)}$ &   biomass formation rate\\
           $\text{q}_\text{{mAb}}$  & $\text{pmol/(cell$\cdot$day)}$ &   secretion  rate of mAb\\
          \hline                    
    \end{tabular}
    \footnotesize{*{cell concentration expressed in molar concentrations \citep{hagrot2019novel}.}}
\end{center}
\end{table}

\begin{figure}
    \begin{center}
    \includegraphics[width = 8.4 cm]{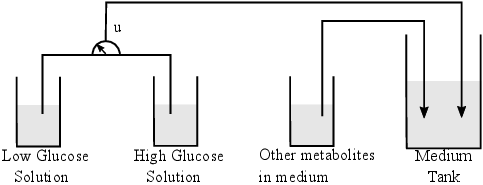}    
    \caption{Schematic of medium with varying glucose concentration.}  
    \label{fig:medium}
    \end{center}
\end{figure}

\subsection{Feedback control of glycosylation}
As stated in Section~\ref{sec:motivating_example}, our goal is to minimize the economic cost induced by $P_{G_0}$  becoming lower than the imposed reference $G0$ glycoform percentage $P_{G_0,opt} = 35\%$. 
 The most straightforward approach is to simply design the medium using available models and relationships, such that $P_{G_0} > P_{G_0,opt}$, e.g.,  \citep{zhang2021control,kappatou2020quality,kotidis2019model}. However, such an approach will be sensitive to model uncertainty and unmeasured disturbances. 
 A straightforward extension is to combine the open-loop design with the proposed learning-based predictive filter to enhance constraint satisfaction under uncertainty. However, because metabolic dynamics are typically slow, the collected data may not be sufficiently informative for learning the optimal constraint margins. To address this, we propose a simple feedback control law integrated with the learning-based predictive filter, demonstrating improved performance under uncertainty compared with open-loop approaches.

We aim to control $P_{G_0}$ by adjusting the glucose concentration in the fed medium, which can be achieved by mixing low ([Glu] = 10 mM) and high ([Glu] = 50 mM) glucose solutions by different percentages, as illustrated in Figure~\ref{fig:medium}. The percentage of low Glucose solution is determined by a proportional controller according to
\begin{equation}
\label{eq:feedback_control_law}
  \small { u =      \begin{cases}
0 \% &\text{if $u < 0\%$}\\
u_0 + k(P_{G_0,opt} - P_{G_0}) \times 100\% &\text{if $u \in [0\%,100\%]$}\\
100 \% &\text{if $u > 100\%$}
\end{cases},}
\end{equation}
where $u$ represents the percentage of low glucose solution; $u_0 = 85.7\%$ is the baseline input value corresponding to a predicted $P_{G_0,opt} = 35\%$ using the available model; $k = 0.3$ is the proportional gain of the controller that adjusts $u$ based on estimates of $P_{G_0}$ with sampling time $T_s = 5 \text{ min}$. The sampling time is motivated by recent advances in monitoring tools such as in-line Raman spectroscopy \citep{schwarz2022monitoring} that allows estimation of the glycosylation profile at steady state based on in-line measurements of the glucose consumption rate within minutes. The open-loop method, on the other hand,  directly applies the medium consisting of  $u_0 = 85.7\%$ of low glucose solution without adjusting the glucose concentration based on the estimates of $P_{G_0}$. Note that the controller (\ref{eq:feedback_control_law}) effectively is a model based feedback controller from in-line measurements of the glucose consumption rate. 

\begin{figure}
    \begin{center}
    \includegraphics[width = 8.4 cm]{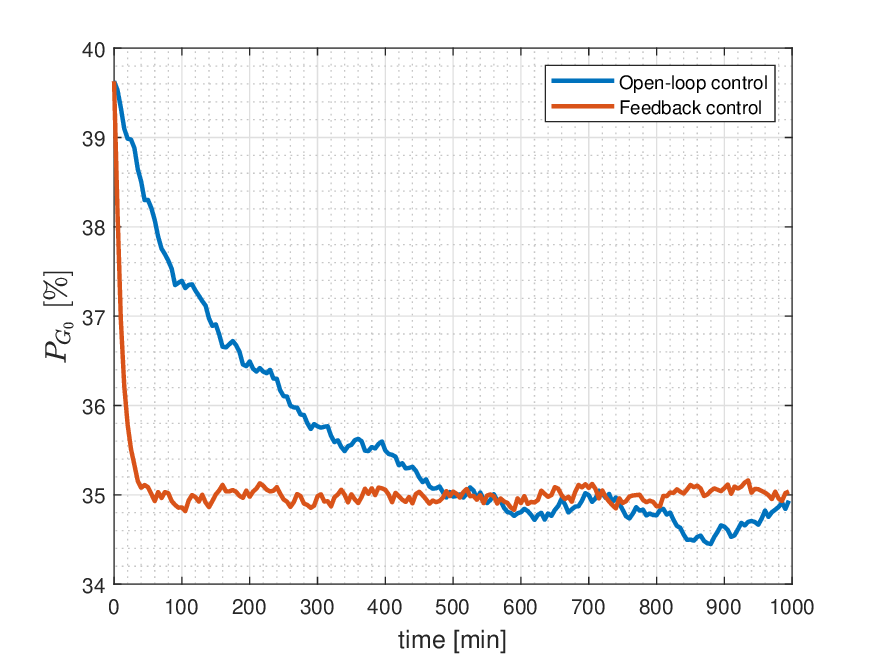}    
    \caption{Comparison of feedback (red) and open-loop (blue) control strategies {with 10\% relative {uniformly distributed} input uncertainty} when regulating $P_{G_0}$ to $35\%$ from a medium with $u = 100\%$ low glucose solution.}  
    \label{fig:open-loop-vs-feedback-no-uncertainty}
    \end{center}
\end{figure}
\begin{figure}
    \begin{center}
    \includegraphics[width = 8.4 cm]{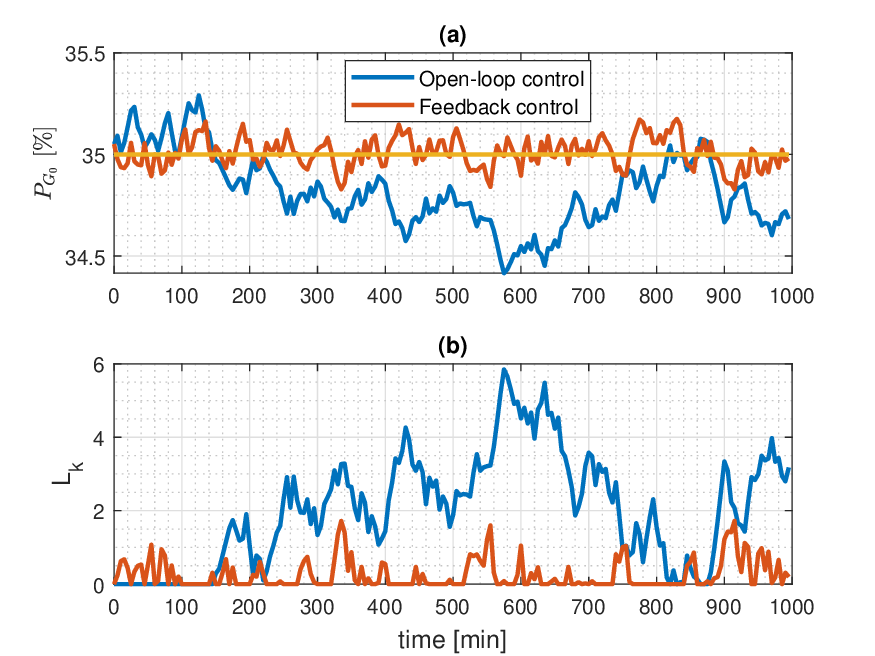}    
    \caption{Comparison of feedback  (red) and open-loop (blue) control strategies with 10\% relative {uniformly distributed} input uncertainty. Fig.  (a) shows the $P_{G_0,opt}$ tracking performance; Fig. (b) shows     constraint violation cost. The solid yellow line represents the optimal operational condition $P_{G_0,opt} = 35\%$.}  
    \label{fig:open-loop-vs-feedback-input-uncertainty}
    \end{center}
\end{figure}

We simulate the system under uniformly distributed input uncertainty of 10\% relative to the nominal control strategy \eqref{eq:feedback_control_law}. Figure~\ref{fig:open-loop-vs-feedback-no-uncertainty} compares the feedback and open-loop methods when regulating the steady-state $P_{G_0}$ estimate to $P_{G_0,opt} = 35\%$ from a medium with $u = 100\%$ low-glucose solution. The feedback controller reaches the target $P_{G_0,opt}$ within $50$ min, whereas the open-loop approach requires more than $500$ min. Figure~\ref{fig:open-loop-vs-feedback-input-uncertainty} further shows that, under the same 10\% input uncertainty, the feedback control maintains robustness and limits constraint violations, while the open-loop method exhibits large deviations and frequent high-cost violations of the constraint $P_{G_0} > P_{G_0,opt}$. The results highlight the importance of feedback control for regulating glycosylation and ensuring consistent product quality under uncertainty. 

We next combine the feedback control strategy with the proposed learning-based predictive filter to further mitigate the economic loss induced by high-cost constraint violations, {under realistic settings}.

\subsection{Constraint handling using learning-based predictive filter}
Consider now applying a learning-based controller that imitates the feedback control law \eqref{eq:feedback_control_law}   to control steady-state $G0$ level glycoform percentage estimate $P_{G_0}$. To account for the explorative behavior of the learning-based controller, we assume the desired input is composed of the proportional control combined with an additive  uniform stochastic part with a magnitude of 5\% of the baseline input value $u_0$, that is, 
\begin{equation}
\label{eq:feedback_learning}
  \small { u =      \begin{cases}
0 \% &\text{if $u < 0\%$}\\
(1+\nu_u)u_0 + k(P_{G_0,opt} - P_{G_0}) \times 100\% &\text{if $u \in [0\%,100\%]$}\\
100 \% &\text{if $u > 100\%$}
\end{cases},}
\end{equation}
 where   $v_{u} \sim U(-0.05,0.05)$ and the other parameters are given in \eqref{eq:feedback_control_law}.   We assume  the $P_{G_0}$  can be estimated accurately by in-line measurements with sampling time $T_s = 5 \text{ min}$. 
\subsubsection{Uncertainties} 
\label{sec:Uncertainties}
{When implementing the desired input signal \eqref{eq:feedback_learning}, we account for potential delays by introducing a time delay ranging from 0 to 0.5 min between the computation of the desired input signal and its actual application to the system, in order to capture delays introduced by experimental devices.}  
We consider the following disturbances
\begin{enumerate}
    \item All states, i.e., metabolites and viable density, subject to uniform disturbances with
    magnitudes that are proportional to viable cell density 
    \small
    \begin{equation}
    \begin{split}
        \frac{d\text{V}[\text{M}_\text{i}]}{d\text{t}} & = \text{q}_{\text{M}_\text{i}}\text{V}[\text{X}_\text{v}] + \text{F}_\text{{in}}[\text{M}_\text{{i,in}}] - \text{F}_\text{{out}}[\text{M}_\text{i}] + d_{\text{M}_\text{i}}, \\
\frac{d\text{V}[\text{mAb}]}{d\text{t}} & = \text{q}_\text{{mAb}}\text{V}[\text{X}_\text{v}] - \text{F}_\text{{in}}[\text{mAb}]+d_{[\text{mAb}]},\\
\frac{d\text{V}[\text{X}_\text{v}]}{d\text{t}} & = \text{q}_\text{{biomass}}\text{V}[\text{X}_\text{v}] - \text{F}_\text{{bleed}}[\text{X}_\text{v}]+d_{[\text{X}_v]},
\end{split}
    \end{equation}
    where
        \begin{equation}
    \small
    \begin{split}
        d_{\text{M}_\text{i}} & = \nu_{d,\text{M}_\text{i}}\text{q}_{\text{M}_\text{i}}\text{V}[\text{X}_\text{v}], \\
d_{[\text{mAb}]} & = \nu_{d,\text{mAb}}\text{q}_\text{{mAb}}\text{V}[\text{X}_\text{v}],\\
d_{[\text{X}_v]} & = \nu_{d,X_v}\text{q}_\text{{biomass}}\text{V}[\text{X}_\text{v}],
\end{split}
    \end{equation}
with $\nu_{d,\text{M}_\text{i}},\nu_{d,\text{mAb}},\nu_{d,X_v} \sim U(-0.05,0.05)$ vary every $0.5$ min.
    \item  to account for other unmeasured disturbances, we introduce an additive disturbance on  $P_{G_0}$ where   the actual $P_{G_0}$ is given by  $P_{G_0}+\nu_{P_{G_0}}$  with $\nu_{P_{G_0}} \sim U(-0.5,0.5)$. 
\end{enumerate}

Note that the size and characteristics of the uncertainty, i.e., disturbances and time delays, are assumed unknown a priori. 

\subsubsection{Results}
In the predictive filter we impose a constraint on the Glucose consumption rate $q_{\mathrm{Glc}}$, which is the measured variable and, furthermore, directly correlated with $P_{G_0}$. This choice is partly motivated by the fact that the dynamics of the glucose consumption rate are well studied, making a predictive model for $q_{\mathrm{Glc}}$ readily available, whereas obtaining a dynamic model for $P_{G_0}$ would require extensive and costly experiments. In the violation cost monitor, the constraint violation cost is evaluated based on a steady-state estimate of $P_{G_0}$ using the glucose consumption rate measurements combined with the static model in \citet{zhang2021control,zhang2020glycan}. 

We hence consider the following constraints in the learning-based predictive filter:
\begin{equation}
    g_h(x,u) = \begin{bmatrix}
    u - 100\%\\
    0\% - u\\
    \end{bmatrix} \leq 0, \,\,\,    g_e(x,u) = q_{Glc} - q_{Glc,ref}
  \leq 0,
\end{equation}
where $g_h(x,u)$ corresponds to the  hard upper and lower constraints on the fraction of low Glucose media, while $g_e(x,u)$ represents high-cost soft constraints with $q_{Glc,ref} = 1.67$ $\text{pmol/(cell$\cdot$day)}$ corresponding to $P_{G_0,opt} = 35\%$ in the absence of uncertainty. The applied penalty function on slack variables in \eqref{eq.predict_filter} is $V_p( \epsilon_{0|k}, \epsilon_{1|k}, \dots, \epsilon_{N|k}) = 10^8 \times \sum_{i=0}^{N} \epsilon_{i|k}$. By adapting the constraint margin  of the high-cost soft constraint  $g_e(x,u)$  in the predictive filter, our objective is to minimize the impact of the uncertainties  on the high-cost constraint violations, i.e., $g_e(x,u) > 0$, with a minimal modification of the desired closed-loop behavior generated by the  feedback controller \eqref{eq:feedback_learning}.  

\begin{figure*}
    \begin{center}
    \includegraphics{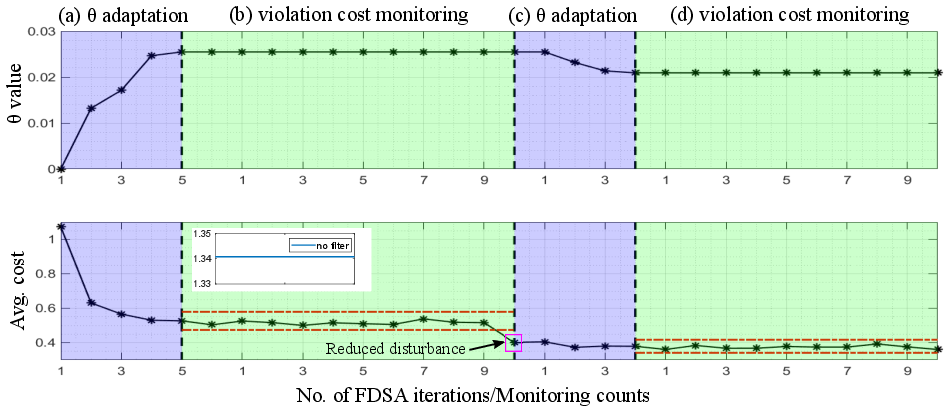}    
    \caption{Constraint adaptation and violation cost monitoring results using the learning-based predictive filter. The  figure above shows the  constraint margin $\theta$ value in the predictive filter, while the figure below shows the corresponding average constraint violation cost obtained by   averaging  \eqref{eq.violate_cost_measure_with_penalty} of 48 samples for each $\theta$. The blue and green regions represent the $\theta$ adaptation and violation cost monitoring phases respectively.}  
    \label{fig:theta_adaptation_water_tank}
    \end{center}
\end{figure*}
\begin{figure}
    \begin{center}
    \includegraphics[width = 8.4 cm]{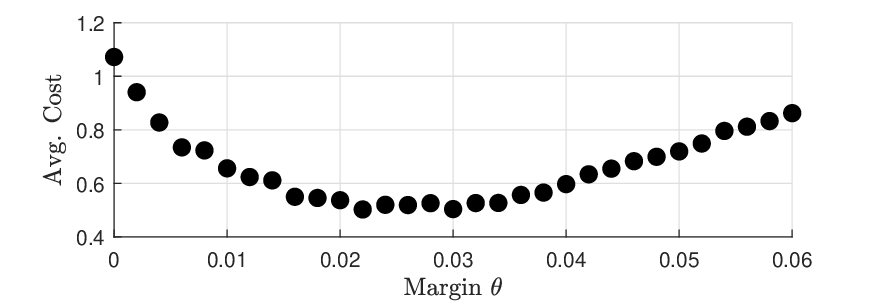}    
    \caption{The relationship between the average constraint violation cost obtained by averaging  \eqref{eq.violate_cost_measure_with_penalty} of 48 samples for each $\theta$ and constraint margin $\theta$. }  
    \label{fig:Fig_theta_cost}
    \end{center}
\end{figure}
\begin{figure}
    \begin{center}
    \includegraphics[width = 8.4 cm]{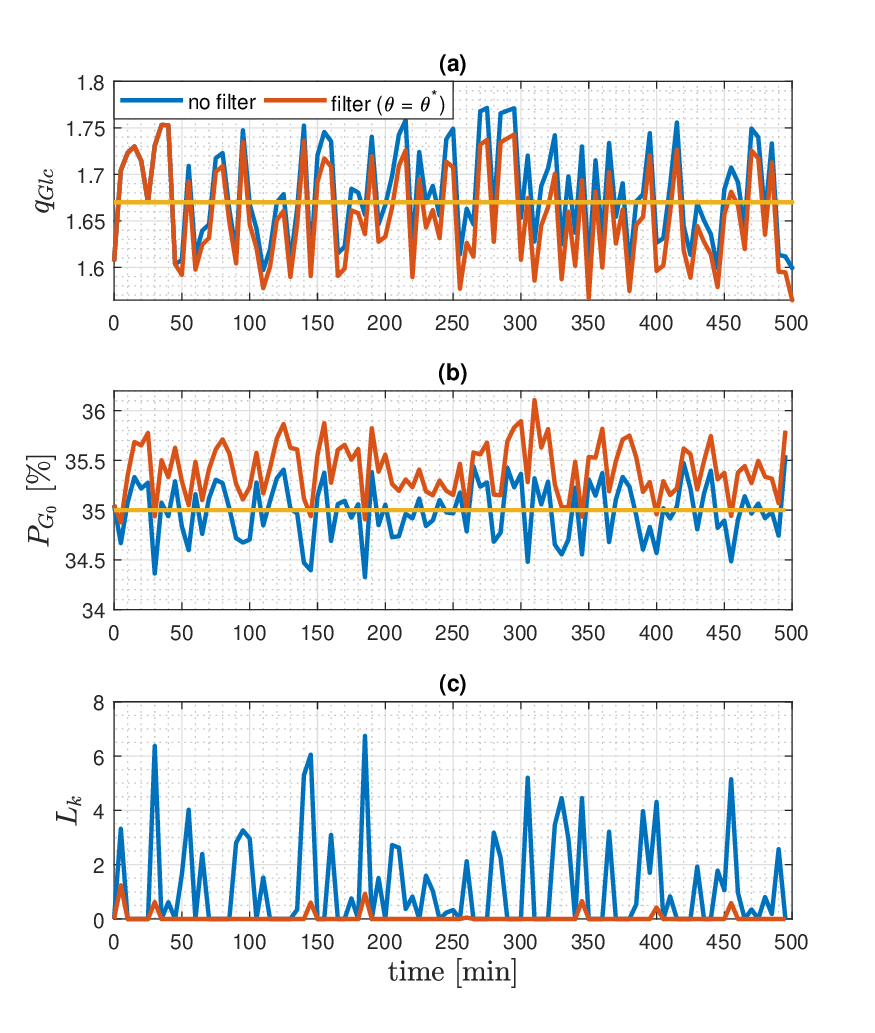}    
    \caption{Comparison of feedback controller \eqref{eq:feedback_learning} with(red)/without(blue) the learning-based predictive filter with the same unknown uncertainties. The predictive filter is with the converged constraint margin $\theta^* =   0.026$.  {Fig.  (a) shows the constraint violation of $q_{Glc}$;} Fig.  (b) shows the $P_{G_0,opt}$ tracking performance;  Fig. (c) shows     constraint violation cost. {The solid yellow lines represent the reference glucose consumption rate $q_{Glc,\text{ref}} = 1.67$~pmol/(cell·day) and the corresponding optimal operational condition, $P_{G_0,\text{opt}} = 35\%$.}}  
    \label{fig:comparison_PI_controller_predictive_filter}
    \end{center}
\end{figure}

To achieve the objective, we apply the learning-based predictive filter described in Section~\ref{sec:adaptive_predictive_filter}. In the constraint adaptor, we  introduce $ 
\theta
\in \mathbb{R}$ as the constraint margin of the high-cost constraints $g_e(x,u)$.  The desired control signal $u^d \in \mathbb{R}$ is generated by the  feedback controller \eqref{eq:feedback_learning}.  In the predictive filter, we apply a discrete-time model obtained by discretizing  the  nominal nonlinear model \eqref{eq:mass_balance_mab} with sampling time $T_s = 5 \text{ min}$. The prediction horizon  is chosen to be $N = 3$. 
 The optimal constraint margin  $\theta$ is learned online  using the FDSA method, with step-size sequences that are adapted from \cite{TROLLBERG20179065} and \cite{spall1998implementation}\footnote{We follow the functional form and parameter choices of \cite{TROLLBERG20179065} with minor rescaling to achieve a practical convergence rate in the simulation setup.}
 \begin{equation*}
     a_j = \frac{1.2\times 10^{-3}}{(j+3)^{0.906}}, \,\,\,\,  c_j = \frac{0.016}{j^{0.4}},
 \end{equation*}
 where $j$ is the number of iterations of FDSA update. The threshold for convergence is chosen to be $\rho_{learning} = 1\times10^{-3}$. 
For each $\theta$ value considered in the gradient approximation step, we  measure the violation cost with penalty term \eqref{eq.violate_cost_measure_with_penalty} where $\lambda = 10$  and $P_k = \max(0,P_{G_0,opt}-P_{G_0})$ by averaging $48$ samples  with the corresponding    $\theta$ value applied in the predictive filter. 
As shown in Figure~\ref{fig:theta_adaptation_water_tank} (a),  $\theta$ quickly converges to $\theta^* = 0.026$  with a low violation cost  within few iterations, starting from  $\theta_{0} = 0$ where the  constraint violation cost is relatively high. The converged  $\theta^* = 0.026$ is in the region with low average violation cost, as illustrated in Figure~\ref{fig:Fig_theta_cost}.  Note that when using the feedback controller \eqref{eq:feedback_learning} without constraint handling (blue line, Figure~\ref{fig:theta_adaptation_water_tank} (b)), the average constraint violation cost is $1.34$, which is more than $2.5$ times larger than the case when using the predictive filter with $\theta = \theta^*$ with the average constraint violation cost  $0.53$. This demonstrates that constraint handling as such is important under uncertainty.

We next compare the 
performance of  the feedback controller \eqref{eq:feedback_learning} with and without the learning-based predictive filter in the presence of the same  uncertainties. As shown in Figure~\ref{fig:comparison_PI_controller_predictive_filter}, the  $G0$  level  glycoform   percentage $P_{G_0}$ is seldom below $P_{G_0,opt} = 35\%$ using the learning-based predictive filter with $\theta^* = 0.026$, while the violation is severe and quite frequent using the feedback controller \eqref{eq:feedback_learning}  only. 
{Note that, due to the minimal modification to the desired input, {the value of $q_{Glc}$ remains close to the desired $q_{Glc}$ when no filter is applied}. Furthermore, the  G0-level glycoform percentage, $P_{G_0}$, achieved using the learning-based predictive filter with the optimal constraint margin $\theta^*$, remains close to the target value $P_{G_0,\text{opt}} = 35\%$, exhibiting similar behavior to the case where only the proportional controller \eqref{eq:feedback_learning} is used.} 
In this example, the desired inputs are frequently modified by the proposed method due to frequent constraint violations under uncertainty. In \cite{wang2022modular}, we presented a fed-batch penicillin production example demonstrating that the desired control input is modified by the proposed method only when necessary. The proposed method ensures that performance is not significantly altered from the desired one while minimizing the impact of high-cost constraint violations under uncertainty. 

When applying the converged optimal  constraint margin $\theta^*$, the violation cost should be monitored constantly such that the constraint margin can be adapted according to the change in the size and characteristics of uncertainties. For this purpose, we apply the proposed constraint violation cost monitor with the number of  samples for averaging $n_{monitor} = 48$ and threshold  $\rho_{monitor} = 10\%$. As illustrated in Figure~\ref{fig:theta_adaptation_water_tank} (b), within the first 10 monitoring counts,   the monitored average constraint violation cost  $J_{monitor}$ falls within the interval $[(1 -\rho_{monitor})J^*,(1 +\rho_{monitor})J^*]$ (red dashed lines) where $J^* = 0.53$ is the average constraint violation cost corresponding to the converged $\theta^* = 0.026$. This implies that the condition   \eqref{eq:monitoring_condition} is satisfied  and the  constraint margin $\theta^*$ is still optimal.  After the $10^{\text{th}}$ monitoring count, we reduce the  size of the disturbances described in Section~\ref{sec:Uncertainties} by $30\%$. As illustrated in Figure~\ref{fig:theta_adaptation_water_tank} (b),  at the $11^{th}$ monitoring count, the monitored cost $J_{monitor}$ falls beyond the red dashed lines, meaning that the monitor detects that  $|{J_{monitor} - J^*}| > \rho_{monitor}|{J^*}|$. This implies that  the current constraint margin $\theta$ may no longer be optimal due to the change in the uncertainties.  Therefore, a new optimal $\theta$ should be learned using the constraint adaptor. As shown in  Figure~\ref{fig:theta_adaptation_water_tank} (c),  a new optimal margin $\theta = 0.021$ is learned, starting from the previous optimal margin $\theta^* = 0.026$, and a lower long-term cost violation cost $0.38$ is reached. After converging to the new constraint margin, the violation cost is constantly monitored again by the violation cost monitor, as illustrated in Figure~\ref{fig:theta_adaptation_water_tank} (d).  This shows that the proposed method  has the ability to adjust the optimal constraint margin in response to the detection of a change in the uncertainties.

\section{Conclusion and Discussion}\label{sec:conclusion}
In this paper we proposed a universal modular approach to constraint handling, based on a predictive filter combined with constraint margin adaptation and constraint violation cost monitoring to minimize the cost of soft constraint violations.  The main advantage of the proposed method is that, as a universal module, it can be easily applied to a system with arbitrary controllers and can ensure constraint satisfaction with minimal modification of the desired closed-loop performance. This provides a simple approach to enable systems to complete tasks using arbitrary controllers or optimizers  without significant economic cost induced by high-cost soft constraint violations. By constantly monitoring the constraint violation cost, the proposed method can adjust optimal constraint margins according to changes in the size and characteristics of uncertainties.  Both the learning of the optimal constraint margin and the detection of the changes in uncertainties are directly from real-time data, and no prior knowledge of the uncertainties is required.  We illustrated the potential of the proposed method using a realistic simulation case study of glycosylation constraint satisfaction in continuous mAb biosimilar production based on a  realistic metabolic network  and glycosylation models involving  23 extracellular metabolites and 126 reactions. The method resulted in a more than 60\% reduction in the cost due to constraint violations as compared to the case without the proposed constraint handling method. The efficiency of the  constraint margin adaptation and constraint satisfaction, while largely maintaining the desired closed-loop behavior, demonstrates the potential of the proposed method for application in 
complex control and optimization systems that do not explicitly handle constraints.

\section*{Acknowledgements}
The authors are grateful to  Dr. H\aa{}kan Hjalmarsson,  Dr. Lukas Hewing  and Dr. Melanie Zeilinger  for valuable feedback, and Dr. Mirko Pasquini for the discussion on the metabolic network model. 

\section*{Funding}
 This work was supported by the VINNOVA Competence Centre AdBIOPRO, diarie nr. 2016-05181 and 2022-03170.
\vspace*{-12pt}

\bibliographystyle{elsarticle-harv}    \biboptions{authoryear}
\bibliography{Manuscript_V2}

\end{document}